# *CryptoRuble: From Russia with Love*

Zura Kakushadze[§¶1] and Jim Kyung-Soo Liew[†‡2]

[§] *Quantigic® Solutions LLC,[3] 1127 High Ridge Road, #135, Stamford, CT 06905*

[¶] *Free University of Tbilisi, Business School & School of Physics
240, David Agmashenebeli Alley, Tbilisi, 0159, Georgia*

[†] *SoKat Consulting, LLC,[4] Woodstock, MD 21163*

[‡] *The Johns Hopkins Carey Business School, 100 International Drive, Baltimore, MD 21202*

November 2, 2017

*"Government's view of the economy could be summed up in a few short phrases:
If it moves, tax it. If it keeps moving, regulate it. And if it stops moving, subsidize it."*
– *Ronald Reagan[5]*

## Abstract

We discuss Russia's underlying motives for issuing its government-backed cryptocurrency, CryptoRuble, and the implications thereof and of other likely-soon-forthcoming government-issued cryptocurrencies to some stakeholders (populace, governments, economy, finance, etc.), existing decentralized cryptocurrencies (such as Bitcoin and Ethereum), as well as the future of the world monetary system (the role of the U.S. therein and a necessity for the U.S. to issue CryptoDollar), including a future algorithmic universal world currency that may also emerge. We further provide a comprehensive list of references on cryptocurrencies.

---

[1] Zura Kakushadze, Ph.D., is the President and CEO and a Co-Founder of Quantigic® Solutions LLC and a Full Professor in the Business School and the School of Physics at Free University of Tbilisi. Email: zura@quantigic.com
[2] Jim Kyung-Soo Liew, Ph.D., is the CEO and Founder of SoKat Consulting, LLC and an Assistant Professor in Finance at the Johns Hopkins Carey Business School. Email: jim@sokat.co and kliew1@jhu.edu
[3] DISCLAIMER: This address is used by the corresponding author for no purpose other than to indicate his professional affiliation as is customary in publications. In particular, the contents of this paper are not intended as an investment, legal, tax or any other such advice, and in no way represent views of Quantigic® Solutions LLC, the website www.quantigic.com or any of their other affiliates.
[4] DISCLAIMER: This address is used by the corresponding author for no purpose other than to indicate his professional affiliation as is customary in publications. The contents of this paper are not intended as an investment, legal, tax or any other such advice, and in no way represents the views of SoKat Consulting, LLC, the website www.SoKat.co or any of their other affiliates.
[5] [GOP, 2017].

*"A cryptocurrency is a digital or virtual currency that uses cryptography for security. A cryptocurrency is difficult to counterfeit because of this security feature. A defining feature of a cryptocurrency, and arguably its most endearing allure, is its organic nature; it is not issued by any central authority, rendering it theoretically immune to government interference or manipulation."* [Investopedia, 2017] A plethora of decentralized cryptocurrencies has emerged since the inception of Bitcoin (BTC) in 2009 following a seminal white paper by Nakamoto [2008],[6] with the total market capitalization now exceeding $170B [CoinMarketCap, 2017].[7]

Decentralization and anonymity of transactions make cryptocurrencies appealing to a range of users,[8] but, unsurprisingly, not to central banks or sovereign governments. However, decentralization and blockchain technology [Nakamoto, 2008] are not synonymous. Blockchain is a ledger for keeping a record of all transactions. Fiat money, albeit imperfectly, can be viewed as an attempt to approximate such recordkeeping[9] [Kocherlakota, 1998].[10] However, there are impeding factors such as fragmented recordkeeping prone to errors and fraud.[11] Even with electronic transactions,[12] notwithstanding the existence of central banking systems, recordkeeping is still local by nature. Blockchain provides a robust (albeit not infallible)[13] technological solution to this problem.[14]

**Government-issued cryptocurrencies?**

Recent reports suggest that Russia will soon issue its government-backed cryptocurrency, ***CryptoRuble*** (see, e.g., [Coldewey, 2017]). Unlike decentralized cryptocurrencies such as Bitcoin and Ethereum, with CryptoRuble (and other expected government-issued cryptocurrencies) there is no mining, all transactions are recorded via blockchain and verified by a centralized government authority. Blockchain (cf. traditional bank ledgers) can help prevent/reduce fraud, errors, etc.

---

[6] For a much earlier work on an anonymous cryptographic electronic money or electronic cash system known as "ecash", see [Chaum, 1983, 1985], which served as a basis for Chaum's electronic money company DigiCash, Inc. founded in 1989. In 1998 DigiCash filed for Chapter 11 bankruptcy and was sold for assets in 2002 [Pitta, 1999].
[7] Cryptocurrency prices are volatile. The $170B figure above is as of October 17, 2017. As of said date, Bitcoin had approximately 55% of the total market share. Table 1 summarizes top 5 cryptocurrencies by market capitalization.
[8] Including those with nefarious intentions, such as money laundering, tax evasion and other illegal activities.
[9] In the U.S. the Bank Secrecy Act of 1970 (or BSA) [Meltzer, 1991], among other things, requires financial institutions to file a Currency Transaction Report (CTR) with the Financial Crimes Enforcement Network (FinCEN) for all transactions exceeding $10,000. The U.S. Government records and stores information about all such transactions internally. This information is invaluable in light of USD being the default reserve currency.
[10] Also see [Kocherlakota and Wallace, 1998]. In the context of Bitcoin, see, e.g., [Luther and Olson, 2015]. Sometimes, the concepts of "ideal money" and "asymptotically ideal money" by Nash [2002] are cited in this context, albeit it is unclear whether there is any direct connection to Bitcoin or cryptocurrencies in general.
[11] A long list of other factors includes irrationality of markets in setting prices, socio-political inequalities, etc.
[12] Which theoretically should be less affected by such issues than physical money.
[13] Mt. Gox's $460M disaster [McMillan, 2014] is only one example of cryptocurrency exchanges being hacked.
[14] See Appendix A for a partial list of literature on Bitcoin, blockchain and cryptocurrencies, and references therein.

However, the aforesaid benefit of using blockchain is only the tip of the iceberg. With decentralized cryptocurrencies, one of the key concerns of central banks and sovereign governments is ceding control. With government-issued cryptocurrencies, central banks and sovereign governments will gain even more control, not less, than with the current banking system. Blockchain maintained by a centralized government authority provides a centralized, un-fragmented ledger of all transactions and other information (such as meta-data) associated therewith. This is any government's dream come true! This will make keeping tabs even easier…

Moreover, as the saying goes, "If you can't beat them, tax them."[15] This is what Russia in part appears to be doing with CryptoRuble. Reportedly (see, e.g., [Coldewey, 2017]),[16] "undocumented" CryptoRubles (those without proof of origin) will be subject to a 13% tax. This is akin to a government-mandated money laundering machine and with such a low overhead should be extremely attractive to all sorts of shady players. Russia will attract not only Russian but also foreign money (including dirty money, but not only). CryptoRuble will likely act as a tax shelter for U.S. and other subjects. One imbedded bonus for the Russian government is that it will own the information encoded in blockchain, including potentially shady transactions, and it can use this information in the future as *Kompromat*[17] against such (unsuspecting) transactors.

However, Russia's (or another at-odds-with-the-West state's) **primary goal** in issuing a government cryptocurrency is to free their monetary system from the controls exerted by the Federal Reserve (Fed), European Central Bank (ECB) and their allied central banks. CryptoRuble creates a buffer layer (see Figure 1) that only the Russian government has control over with pertinent information inaccessible to the U.S., the E.U., etc. Russian elite/oligarchs can launder their money using CryptoRuble, become impervious to (or less affected by) economic sanctions, make their assets currently tied up because of the U.S./E.U. controls (more) liquid, and so on.

---

[15] A parallel with an argument for legalizing cannabis/other illicit drugs and regulating and taxing them (as opposed to fighting them) can be drawn here. We should stress that we are not advocating for or against such legalization here, nor are we making an argument for or against state-mandated money laundering. We merely draw a parallel. Another, somewhat different parallel can be drawn with taxing disruptive businesses, e.g., Uber [McArdle, 2016].
[16] The original report appeared on October 14, 2017 in a Russian newspaper *"Argumenti i Fakti"* [AiF, 2017].
[17] In Russian, *Kompromat* is a shorthand for "compromising material" and stands for damaging information that may be used to blackmail or coerce someone. (There are many such shorthand words in the Russian language.)

For instance, consider an oligarch targeted by sanctions. With the advent of CryptoRuble this oligarch could setup a new shell company (e.g., via a convoluted web of trusts) to continue a prior line of business (e.g., exports of Russian oil or some other commodity). Revenues of this shell company can be legally converted into CryptoRubles. These CryptoRubles can then be converted into funds (or property, good/services, etc.) accessible by the oligarch. Since CryptoRuble uses a centralized blockchain as a non-distributed ledger only available to the Russian government, the conversion of the CryptoRubles into the funds accessible by the oligarch remains hidden from the Fed/ECB/U.S./E.U. The Russian government may tax these funds at 13% if the oligarch does not disclose their origin, but this is a small price to pay. Thus, CryptoRuble will allow various individuals and companies to skirt sanctions. Reportedly, North Korea is already suspected of using Bitcoin for precisely this purpose [Boylan and Taylor 2017].

Undoubtedly, CryptoRuble will disrupt the current world monetary system, the Fed, ECB, U.S. and E.U. policies, their law enforcement operations, etc. Drug money will be easier to clean. Anti-money laundering (AML) controls and efforts will be adversely affected – with the anticipated 13% haircut, this Russian government-sponsored money laundry is simply too attractive to those with nefarious intents and will undoubtedly incent shady players.[18] Money will flow away from the U.S./E.U./U.K. and into Russia. Other countries will likely issue their own cryptocurrencies. Other BRICS members (Brazil, China, India and South Africa)[19] are natural candidates. Reportedly, China is also considering issuing its own cryptocurrency [Yanfei, 2017]. Other countries that could jump on the bandwagon might be Switzerland (considering its history of banking secrecy largely dismantled by the U.S. efforts), Baltic states (which tend to be forward-thinking), Nordic states, Singapore, and perhaps some others. Reportedly, even Spain is considering creating a national blockchain-based ledger [Shieber, 2017], although it is unclear what impact this might have for the (E.U.) monetary system as Spain is part of the Eurozone.

---

[18] Multiple government-issued cryptocurrencies (see below) will only exacerbate deterioration of AML controls.
[19] Reportedly, India is considering introducing government-issued cryptocurrency "Lakshmi" [Bhayani, 2017]. Dubai is contemplating government-issued cryptocurrency "emCash" [Dubai DED, 2017]. Estonia [Browne, 2017a] and Kazakhstan [Browne, 2017b] are also looking into government-issued cryptocurrencies. Japanese banks are eyeing their own digital currency called "J-Coin" (to be pegged with JPY) [Kharpal, 2017]. And the list is ever growing.

A key difference between government-issued and decentralized cryptocurrencies is that the former do not require mining. Mining is a process whereby GPUs/CPUs solve mathematical problems by a brute-force method to verify transactions. Miners are rewarded with small payments in respective cryptocurrencies. One issue with mining is that it is computationally and energy costly and has been argued to be inefficient and possibly even unsustainable (see, e.g., [Bariviera, 2017], [Kostakis and Giotitsas, 2014], [Nadarajah and Chu, 2017], [Urquhart, 2016], [Vranken, 2017]) as blockchain grows with each verified transaction. Centralized government-issued cryptocurrencies do not require mining, in some sense are more "efficient", should be cheaper to maintain and more readily usable as legal tender in large economies such as Russia.

The new era is being ushered in. The world order as we know it is changing, right before our eyes. This disruptive technology – cryptocurrencies – will indeed end up disrupting the status quo. However, at least in the mid-term, forward-thinking sovereign states that embrace and adapt it to their advantage will end up being the disruptors as opposed to disrupted. The U.S. is the sovereign state with most to lose in this process, with a clear policy implication: adapt to the changing reality, ***issue CryptoDollar now***, or risk being marginalized.[20]

**Implications to stakeholders**

What about other stakeholders? In Table 2 we outline some stakeholders with pros and cons from government-issued cryptocurrencies. E.g., the populace benefits from lower transaction costs, but the price to pay is diminished privacy and a greater control by the government.

As another example consider banks and other traditional financial institutions. Government-issued cryptocurrencies will diminish some functions of, e.g., banks since the traditional local-by-nature bank ledgers will become obsolete. So will paper money with the implications to ATMs and bank fees associated with them. Generally, transaction costs will be reduced, a bad news for banks. However, this does not make banks obsolete as they have a number of other functions, e.g., provision of credit (credit cards, mortgages, car loans, etc.).

CryptoRuble will have implications for existing decentralized cryptocurrencies such as Bitcoin and Ethereum, etc. In the short term, higher volatility for decentralized cryptocurrencies can be expected due to two competing perceptions: i) government-issued cryptocurrencies go against the key premise of decentralized cryptocurrencies, to wit, decentralization; and ii) on the other hand, they lend increased credibility to blockchain technology. In the beginning there likely will be a lot of noise and uncertainty, political jostling, security concerns, fear-mongering in press coverage, etc. However, it is feasible that in the longer-term decentralized cryptocurrencies will benefit from increased credibility and even more users will jump on the bandwagon, both those motivated by a major world power issuing its own cryptocurrency as well as those perceiving this as sovereign governments attempting to exert more control. Interesting times lie ahead.

---

[20] For CryptoDollar to be competitive, reducing tax rates and not taxing non-US-derived income might be required.

**iCurrency?**

It is likely not a stretch to imagine that 100 years from now there will be no cash or paper money.[21] Just as bronze, silver and gold coins, paper money will soon seem archaic if not barbaric. One may argue that currently used electronic transactions already are replacing cash. However, the current banking system itself is based on an outdated technology. Will blockchain replace it? It is conceivable that other technologies will replace blockchain. Only time will tell.

Introduction of government-issued cryptocurrencies is a pivotal moment for humankind. While such cryptocurrencies do not meet the four criteria we set forth in [Kakushadze and Liew, 2015a] for *iCurrency*, a purely algorithmic universal numeraire and possibly a universal world currency, in the long-term the rise of government-issued cryptocurrencies may well seed the process ultimately leading to iCurrency. Assuming a bright future for humankind devoid of wars, with a truly globalized economy and interplanetary travel with colonies on Mars, it is difficult to imagine that kind of a world without a universal world currency. iCurrency is defined as a universal numeraire subject to the following criteria [Kakushadze and Liew, 2015a]:

1. iCurrency is not a currency issued (or backed) by any government;
2. it is valued based solely on supply and demand;
3. it is easily transferred across regions and globally accepted as a payment method; and
4. it is algorithmic, with no human intervention.

---

[21] Science fiction, for what this is worth, usually predicts some kind of universal "credits", purely electronic in nature. See, e.g., [Ebert, 1999], [Gliddon, 2005].

It is conceivable that a broad consortium of sovereign governments issues iCurrency, in which case Criterion 1 above can still be preserved so long as robust mechanisms are set in place to prevent any government or a group of governments from engaging in possible manipulation of iCurrency, e.g., via (analogs of) central banks artificially increasing or decreasing interest rates, monetary supply manipulation, political influences, and so forth. Criterion 2 above may be most challenging in the context of achieving stability and low volatility.[22] In this regard, a staged rollout of iCurrency may be warranted, whereby its value is determined by a combination of supply and demand and regulation by a central authority. E.g., the USD/HKD spread trades between 7.75 and 7.85, a band fixed by the Hong Kong Monetary Authority. iCurrency too could be regulated at first to stay in such a target zone, which could be gradually relaxed and perhaps removed entirely at some future time. Or perhaps some wide target zone could remain.[23]

iCurrency would allow to more precisely estimate global inflation, global output, global productivity, global labor gains, and other global macroeconomic indicators, as well as GDP (gross domestic product) for various countries and regions and globally. (GDP measured using iCurrency was termed *iGDP* in [Kakushadze and Liew, 2015a]). With low sovereign government intervention and low fees for global transactions, iCurrency should be a boon to our global economy. Perhaps somewhat ironically, government-issued cryptocurrencies may be a step forward (not backward) in this regard, depending on how the world leaders approach this issue.

At first, it may be unexpected that Russia, out of all the countries, has emerged as a "leader" in this process. However, shortly after we published our paper [Kakushadze and Liew, 2015a], it became evident to us that Russia was in a unique position to lead this process [Kakushadze and Liew, 2015b].[24] And the reason is not technological but geopolitical. Russia has been subject to sanctions and global pressure due to its various policies w.r.t. its immediate neighbors and elsewhere in the world. Dependence on the existing world monetary order is a major stumbling block for Russia. Its desire to gain a greater degree of independence from the Fed/ECB is by no means a surprise. Cryptocurrencies arguably provide a simple solution for

---

[22] Generally, government-issued cryptocurrencies are backed by taxation power and expected to be less volatile.
[23] In which case iCurrency would not be 100% based on supply and demand, but could be close to it for all practical purposes and the band would ensure some nominal degree of stability in case of unforeseen impactful events.
[24] We should have published [Kakushadze and Liew, 2015b] back then. But hindsight is always 20-20…

Russia. One could imagine that, if pushed into a corner, Russia could even adopt Bitcoin as a legal tender [Kakushadze and Liew, 2015b]. However, that would mean ceding too much control over information. Somewhat less dramatically, the next natural move for Russia is to issue its own government-backed cryptocurrency, CryptoRuble. If anything, it is surprising it took them so long.

It appears that government-issued cryptocurrencies are a foregone conclusion. Large sovereign states have the technological know-how and means to do this. What about small and/or developing countries? If they are forced to outsource issuance of their government-backed cryptocurrencies to larger states, geopolitical and economic implications are evident: less sovereignty and ceding control over crucial information to more powerful countries.[25]

---

[25] Alternatively, some countries could partner with the private sector, as appears to be the case with Dubai [Dubai DED, 2017]. This would likely raise its own host of issues.

arrangements. *Journal of Economic Theory* **81**(2): 272-289.

Kostakis, V. and Giotitsas, C. (2014) The (A)Political Economy of Bitcoin. *Journal for a Global Sustainable Information Society* **12**(2): 431-440.

Luther, W.J. and Olson, J. (2015) Bitcoin is Memory. *Journal of Prices & Markets* **3**(3): 22-33.

McArdle, M. (2016) Robbing the New (Uber) to Subsidize the Old (Taxis). *Bloomberg View* (August 23, 2016). https://www.bloomberg.com/view/articles/2016-08-23/massachusetts-tax-robs-from-uber-to-subsidize-taxis.

McMillan, R. (2014) The Inside Story of Mt. Gox, Bitcoin's $460 Million Disaster. *Wired* (March 3, 2014). https://www.wired.com/2014/03/bitcoin-exchange/.

Meltzer, P.E. (1991) Keeping Drug Money from Reaching the Wash Cycle: A Guide to the Bank Secrecy Act. *Banking Law Journal* **108**(3): 230-255.

Nadarajah, S. and Chu, J. (2017) On the inefficiency of Bitcoin. *Economics Letters* **150**: 6-9.

Nakamoto, S. (2008) Bitcoin: A Peer-to-Peer Electronic Cash System. *Working Paper.* https://bitcoin.org/bitcoin.pdf.

Nash, Jr., J.F. (2002) Ideal Money. *Southern Economic Journal* **69**(1): 4-11.

Pitta, J. (1999) Requiem for a Bright Idea. *Forbes* (November 1, 1999). https://www.forbes.com/forbes/1999/1101/6411390a.html.

Shieber, J. (2017) Spain's biggest companies are charging into crypto. *TechCrunch* (October 17, 2017). https://techcrunch.com/2017/10/17/spains-biggest-companies-are-charging-into-crypto/?ncid=rss.

Urquhart, A. (2016) The inefficiency of Bitcoin. *Economics Letters* **148**: 80-82.

Vranken, H. (2017) Sustainability of Bitcoin and blockchains. *Current Opinion in Environmental Sustainability* **28**: 1-9.

Yanfei, W. (2017) PBOC inches closer to digital currency. *China Daily* (October 14, 2017). http://usa.chinadaily.com.cn/business/2017-10/14/content_33236132.htm.

## Appendix A: Further Reading

The literature cited below constitutes a partial list, mostly (but with some exceptions) consists of peer-reviewed articles, and intentionally avoids ubiquitous conference proceedings articles. There are other (mostly unpublished) papers available from SSRN, arxiv.org and other sources.

| Rank | Name | Symbol | Market Cap, $B | Price, $ | Volume (24h), $M |
|---|---|---|---|---|---|
| 1 | Bitcoin | BTC | 94.165 | 5662.60 | 1,817 |
| 2 | Ethereum | ETH | 30.484 | 320.36 | 459 |
| 3 | Ripple | XRP | 9.374 | 0.243270 | 518 |
| 4 | Bitcoin Cash | BCH | 6.045 | 361.91 | 854 |
| 5 | Litecoin | LTC | 3.223 | 60.34 | 199 |

**Table 1.** Top 5 cryptocurrencies by market capitalization. The data (market capitalization and volume are rounded) is taken from [CoinMarketCap, 2017] as of approximately 1:00 PM EDT on October 17, 2017.

| Stakeholder | Pros | Cons |
|---|---|---|
| Populace | Low transaction costs | Security and privacy concerns, government control |
| Issuing governments | Control over transactions, information | Cannibalization of the existing monetary system |
| Other governments | Loss of control (for some) | Lower regulatory burdens |
| Banks | Low error and fraud rates | Lower fees |
| Small Businesses (including FinTech, but not only) | Higher growth, increased opportunities for new technological developments | Reputational risks (money laundering, fraud, etc.) |
| Finance | Higher growth, new asset classes | Destabilization of the current business models |
| Economy | Higher growth, facilitated internet commerce | Learning curve, reputational risks |
| Non-government organizations (NGOs) | Injecting technology into developing countries | Learning curve |
| Heads of states | Forward-looking/progressive thought leadership, control over information and monetary transactions | Resistance from opposition (to authoritarian tendencies, etc.) |
| Marketplace participants | Simpler/more transparent monetary transactions | Increased government control and potential interference |
| Decentralized cryptocurrencies (BTC, ETH, etc.) | Creating perception of legitimacy (longer-term) | Even higher volatility then currently (short-term) |

**Table 2.** Potential pros and cons to some stakeholders from government-issued cryptocurrencies.

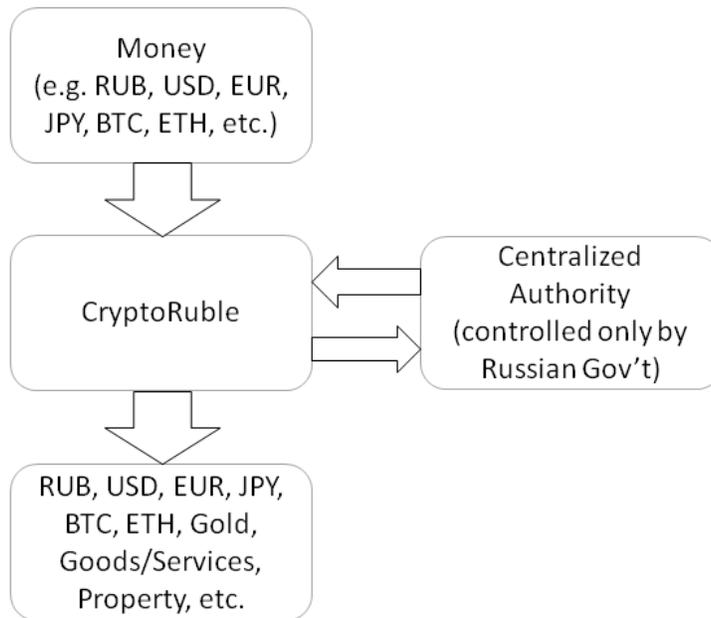

**Figure 1.** A schematic depiction of the money flow into and from CryptoRuble. CryptoRuble creates a buffer layer that only the Russian government has control over with pertinent information inaccessible to the Fed/ECB (or their allied central banks), the U.S., the E.U., etc.